\documentclass{article}
\usepackage{acronym}
\acrodef{3GPP}{3rd generation partnership project}
\acrodef{A2A}{all-to-all}
\acrodef{A2AV}{all-to-all-v}
\acrodef{AI}{artificial intelligence}

\acrodef{CDF}{cumulative distribution function}
\acrodef{DiT}{Diffusion Transformer}
\acrodef{DA}{domain adaptation}
\acrodef{DG}{domain generalization}
\acrodef{DL}{deep learning}
\acrodef{DNN}{deep neural network}
\acrodef{DQN}{deep Q-network}
\acrodef{DRL}{deep reinforcement learning}
\acrodef{EP}{expert parallelism}
\acrodef{FFN}{feed forward network}
\acrodef{GPU}{graphics processing unit}
\acrodef{GAN}{generative adversarial network}
\acrodef{GP}{gaussian process}
\acrodef{HBM}{high-bandwidth memory}
\acrodef{IoT}{internet of things}
\acrodef{IPC}{inter-process communication}
\acrodef{IB}{InfiniBand}
\acrodef{KL}{Kullback-Leibler}
\acrodef{KPI}{key performance indicator}
\acrodef{LLM}{large language model}
\acrodef{MoE}{mixture-of-expert}
\acrodef{MAE}{mean absolute error}
\acrodef{MDP}{Markov decision process}
\acrodef{ML}{machine learning}
\acrodef{MIMO}{multiple-input and multiple-output}
\acrodef{MLP}{multi-layer perceptron}
\acrodef{NPU}{neural processing unit}

\acrodef{PDF}{probability density function}
\acrodef{PPO}{proximal policy optimization}
\acrodef{QoE}{quality of experience}
\acrodef{QoS}{quality of service}
\acrodef{RDMA}{remote direct memory access}
\acrodef{RL}{reinforcement learning}

\acrodef{TTFT}{time to first token}
\acrodef{TPOT}{time per output token}
\acrodef{UB}{unified buffer}

\usepackage[preprint]{aaj2026}
\usepackage[utf8]{inputenc}
\usepackage[T1]{fontenc}
\usepackage{hyperref}
\usepackage{url}
\usepackage{booktabs}
\usepackage{amsfonts}
\usepackage{amsmath}
\usepackage{amssymb}
\usepackage{nicefrac}
\usepackage{microtype}
\usepackage{xcolor}
\usepackage{graphicx}
\usepackage{multirow}
\usepackage{enumitem}
\usepackage{array}
\usepackage{subcaption}
\usepackage{svg}
\usepackage{algorithm}
\usepackage{algpseudocode}

\title{Relay Buffer Independent Communication over Pooled HBM for Efficient MoE Inference on Ascend}
\author{%
\begin{tabular}{c}
Tianlun Hu, Tiancheng Hu, Shengsheng Litang, Sheng Wang, Xiaoming Bao, Yuxing Li\\
Wei Wang, Zhongzhe Hu, Lijun Li, Hongwei Sun, Jingbin Zhou\\
\end{tabular}
\\
Huawei Technologies 
}

\begin{document}

\maketitle
\begin{abstract}
\Ac{MoE} inference relies on large-scale token exchange across devices, making dispatch and combine major bottlenecks for both throughput-oriented prefill and latency-sensitive decode. Prior work has shown that the cost of \ac{MoE} communication is not determined by network transfer alone: routing-driven layout transformation, temporary relay, and output restoration can contribute a substantial fraction of the end-to-end overhead. However, many existing \ac{MoE} communication paths remain fundamentally buffer-centric, relying on explicit \ac{IPC} relay and intermediate reordering buffers before or after collective transfer. In this report, we present a relay-buffer-free communication design for \ac{MoE} inference acceleration on Ascend systems. Rather than treating dispatch and combine as a conventional relay-and-restore procedure, the proposed design reorganizes them around direct placement into destination expert windows and direct reading from remote expert windows. Built on top of globally pooled \ac{HBM} and symmetric-memory allocation, the proposed implementation reduces explicit intermediate relay and reordering buffers while retaining only lightweight control state such as counts, offsets, and synchronization metadata. We further instantiate this design as two schedules tailored to the two main phases of \ac{MoE} inference. The prefill schedule preserves richer planning state to support throughput-oriented execution, while the decode schedule uses a more compact procedure to reduce latency-sensitive overhead. Experimental results on Ascend-based \ac{MoE} workloads show that the proposed implementation reduces dispatch and combine latency across both prefill and decode settings. At the end-to-end serving level, it improves \ac{TTFT} while preserving competitive \ac{TPOT}, and enlarges the feasible scheduling space under practical latency constraints. These results suggest that, on platforms with globally addressable device memory, reducing intermediate buffering and output restoration around expert execution is an effective direction for accelerating \ac{MoE} inference.
\end{abstract}

\section{Introduction}

\Ac{MoE} has become a major architectural direction for scaling modern foundation models. By activating only a small subset of experts for each token, \ac{MoE} substantially increases model capacity without proportionally increasing per-token computation, and has been widely adopted in modern \acp{LLM}, industrial-scale sparse models, and increasingly in multi-modal systems \cite{shazeer2017outrageously,fedus2022switch,rajbhandari2022deepspeedmoe,hwang2023tutel,gale2023megablocks,deepseek2024v3}. This trend reflects a broader shift in large-scale model design: instead of scaling dense computation uniformly, modern systems increasingly rely on sparse activation to improve the capacity-efficiency trade-off.

A major challenge in \ac{MoE} inference is inter-device token exchange. Under \ac{EP}, experts are partitioned across devices, and routed tokens must be dispatched to remote experts and later combined back into the original logical order. As \ac{EP} scales, this distributed shuffling overhead increasingly becomes a system bottleneck. Prior work has shown that \ac{MoE} execution efficiency is highly sensitive to routing-induced dynamic workload imbalance, communication-computation coordination, and the way communication is organized around expert execution \cite{hwang2023tutel,rajbhandari2022deepspeedmoe,deepseek2024v3}. In practice, the bottleneck is not merely network transfer itself, but the full execution path surrounding token movement, including packing, permutation, staging, synchronization, and layout restoration.

This problem becomes more critical as foundation models continue to push toward much longer contexts. DeepSeek-V3 reports strong performance at 128$K$ context length \cite{deepseek2024v3}. Qwen2.5-1M extends context to 1$M$ tokens and relies on techniques such as chunked prefill, sparse attention, pipeline parallelism, and scheduling optimization to make long-context inference practical \cite{yang2025qwen25_1m}. In the multimodal setting, Qwen2.5-VL targets long-video understanding with videos of up to hours in duration \cite{wang2025qwen25_vl}. Taken together, these trends place increasing pressure on memory systems, transient buffers, and communication paths.

Unfortunately, many existing \ac{MoE} communication paths remain fundamentally buffer-centric. In a conventional implementation, tokens are first packed or permuted into communication-friendly layouts, transferred through generic collectives such as \ac{A2A} or \ac{A2AV}, and then restored into expert-major or token-major layouts for subsequent computation. This mechanism introduces extra memory traffic and transient buffer footprint, and it sets a rigid separation between communication and operator execution. The limitations of such formulations have already been reflected in prior \ac{MoE} systems work. Tutel emphasizes dynamically adaptive parallelism and pipelining under dynamic token routing \cite{hwang2023tutel}, while DeepEP provides \ac{MoE}-specific high-throughput and low-latency all-to-all kernels for dispatch and combine rather than relying solely on generic collectives \cite{deepep}. Recent work such as FUSCO further shows that pre- and post-communication rearrangement can account for a substantial fraction of total shuffle time \cite{fusco2026}.

In addition to latency overhead, buffer-centric execution also increases transient \ac{HBM} footprint. Explicit staging, \ac{IPC} relay, and temporary reordering buffers consume memory that could otherwise be used to increase serving batch size or to accommodate longer contexts. This issue is especially relevant for production inference, where decode steps are latency-sensitive and memory headroom is often limited. Long-context inference engines already rely on chunked prefill and related scheduling techniques, further highlighting the pressure on the underlying communication and memory path \cite{yang2025qwen25_1m}.

In this work, we focus on improving \ac{MoE} inference by reducing not only communication bandwidth cost, but also the buffer-centric data movement around dispatch and combine. Our target platform, Ascend, provides a practical opportunity for this direction. With globally pooled \ac{HBM}, the communication domain can expose device memory regions across ranks, enabling data exchange beyond the conventional sender-pack/receiver-unpack model. In our implementation, this capability is exposed through symmetric-memory allocation and shmem-style remote access, and practical deployment experience suggests that Ascend devices benefit from read-favored execution. These properties motivate a communication mechanism that minimizes explicit intermediate buffering and executes dispatch and combine directly over globally pooled \ac{HBM}.

We therefore present a relay-buffer-free communication design for accelerating \ac{MoE} dispatch and combine on Ascend. The proposed implementation reduces explicit \ac{IPC}-buffer-based relay and reordering by organizing dispatch and combine around destination expert windows and remote expert windows over globally pooled \ac{HBM}. This design follows a hardware-aware, read-favored execution rule and targets the two most communication-sensitive stages in sparse expert-parallel inference. Overall, this report treats \ac{MoE} inference bottlenecks not only as a collective-bandwidth problem, but also as a buffer-centric distributed-execution problem around dispatch and combine.

This report describes the following technical elements:
\begin{itemize}[topsep=4pt,itemsep=3pt,parsep=0pt,partopsep=0pt]
    \item We present a relay-buffer-free communication design for \ac{MoE} inference acceleration that leverages globally pooled \ac{HBM} on Ascend to reduce explicit \ac{IPC}-buffer-based relay and reordering around dispatch and combine.
    \item We describe a hardware-aware, read-favored communication kernel execution routine for Ascend and show how it supports direct placement and direct reading in DeepEP-style \ac{MoE} communication operators.
    \item We instantiate the proposed design for \ac{MoE} dispatch and combine in both prefill and decode, reducing extra copies and simplifying offset handling in sparse expert-parallel inference.
    \item We evaluate the proposed implementation on Ascend systems and report improvements in kernel-level latency and end-to-end \ac{MoE} serving performance across practical use cases.
\end{itemize}

\begin{figure}[t]
    \centering
    \includegraphics[width=0.8\linewidth]{"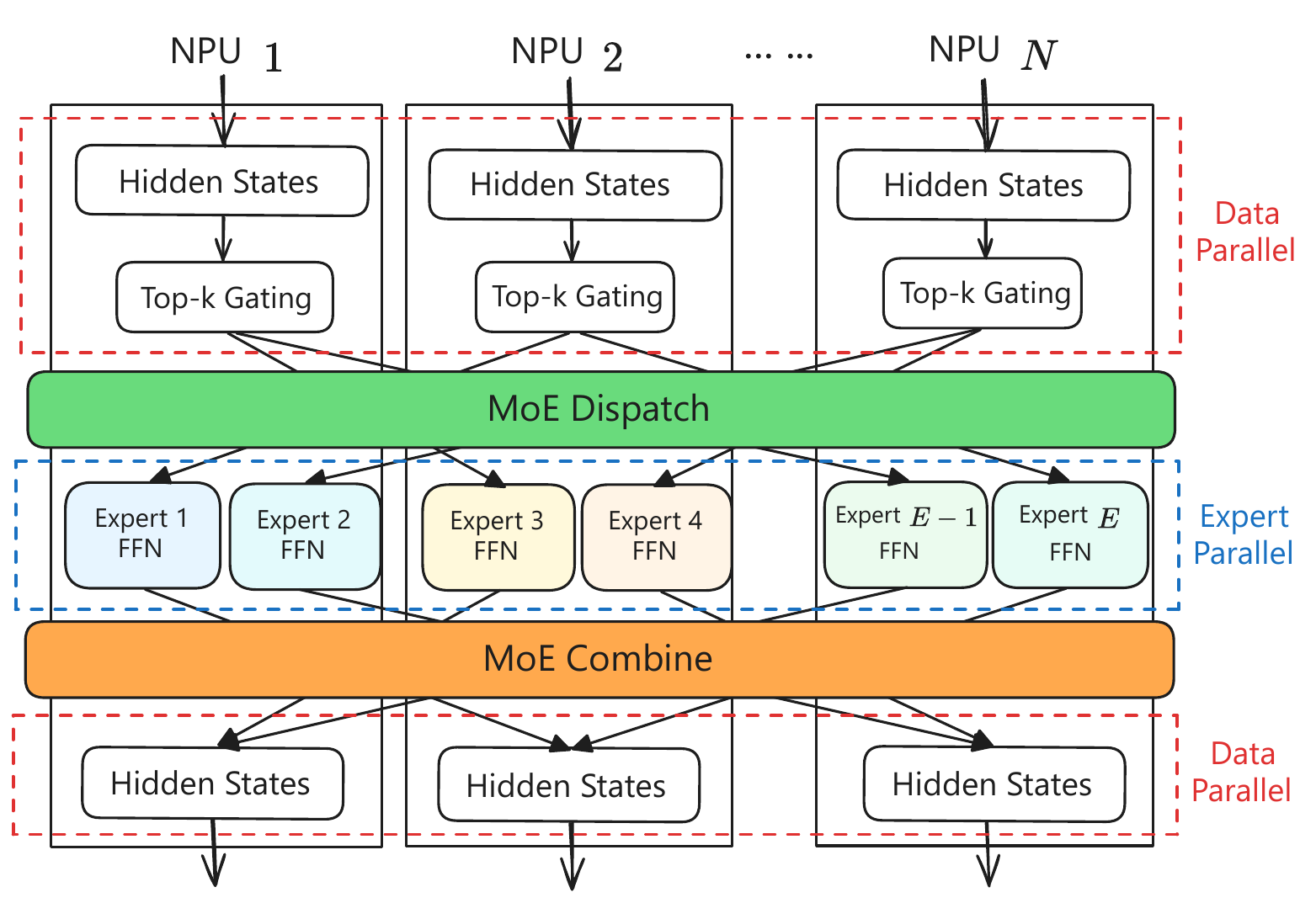"}
    \caption{Conceptual illustration of the position of dispatch and combine in an \ac{MoE} layer. Gating is performed in the data-parallel stage to determine the destination experts for each token. Dispatch redistributes tokens from the data-parallel hidden-state representation to expert-parallel \acp{FFN}, and combine gathers expert outputs back to the data-parallel hidden-state representation for the subsequent layer. This report focuses on accelerating the dispatch and combine stages.}
    \label{fig:moe-dispatch-combine-overview}
\end{figure}

\section{Background and Motivation}

In \ac{MoE} inference, routed tokens must be sent to the devices hosting the target experts and the resulting expert outputs must then be gathered back to the hidden-state order expected by subsequent layers. Under \ac{EP}, this creates two communication-dominated stages:

\textbf{dispatch}, which places routed tokens into expert-side buffers, and \textbf{combine}, which gathers expert outputs and restores the target output order.

The difficulty is not only inter-device transfer. Token-to-expert mapping is irregular, expert loads change at runtime, and the layouts preferred by communication and expert computation are often different. Conventional implementations therefore decompose dispatch and combine into packing, transfer, and restore, introducing explicit staging or reorder buffers along the way. We refer to this pattern as \textbf{buffer-centric execution}. It adds extra \ac{HBM} traffic, increases temporary-memory pressure, and lengthens the latency-critical dispatch/combine path. Prior \ac{MoE} systems have highlighted the same issue from different angles, including DeepSpeed-MoE, Tutel, BigMac, and FUSCO \cite{rajbhandari2022deepspeedmoe,hwang2023tutel,jin2025bigmac,fusco2026}. In particular, FUSCO decomposes \ac{MoE} shuffle into preprocessing, rearrangement, and communication, showing that layout transformation around communication can itself become a major part of shuffle overhead. Figure~\ref{fig:moe-dispatch-combine-overview} illustrates the role of dispatch and combine in the \ac{MoE} execution path.

This cost is especially visible in modern serving workloads. Long-context and multi-modal inference increase both communication volume and transient-memory pressure \cite{cai2024moesurvey,deepseek2024v3,yang2025qwen25_1m,wang2025qwen25_vl}, while decode leaves little room to hide packing, offset translation, staging, and restore. As a result, production \ac{MoE} systems increasingly use dispatch/combine-specific communication paths rather than treating these stages as ordinary collectives \cite{deepep,deepseek2024v3}. Recent production-oriented implementations also move toward platform-specific one-sided communication; for example, TensorRT-LLM includes an NVLink one-sided \ac{A2A} path based on symmetric memory for \ac{MoE} communication within an NVLink domain \cite{tensorrtllm_onesided}.

This work is motivated by an existing Ascend capability: globally pooled \ac{HBM} exposed through symmetric-memory allocation and shmem-style remote access. We leverage this pooled-memory capability to reorganize dispatch and combine around \emph{direct placement into destination expert windows} and \emph{direct reading from remote expert windows}, rather than sender-pack / relay / receiver-restore pipelines.

\begin{figure}[t]
\centering
\includegraphics[width=0.8\linewidth]{"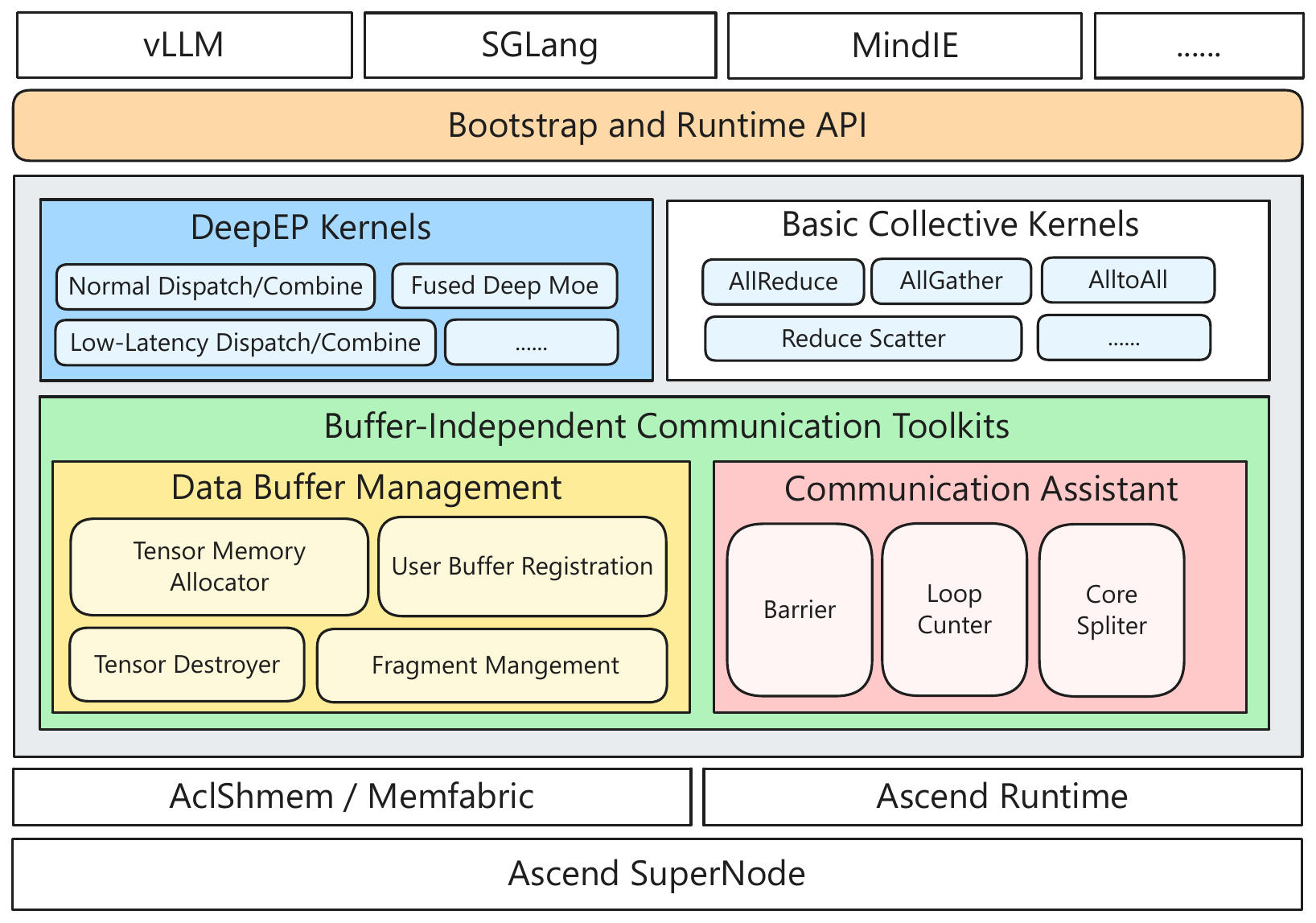"}
\caption{Overview of the proposed relay-buffer-free \ac{MoE} communication path on Ascend. The implementation sits between upper-layer serving frameworks and the globally pooled \ac{HBM} exposed through AclShmem/Memfabric. The system includes bootstrap and runtime APIs, specialized kernels for normal and low-latency dispatch/combine, basic collective kernels, and runtime components for buffer management and communication coordination.}
\label{fig:zerobufferep-overview}
\end{figure}
\section{Design of the Relay-Buffer-Free Communication Path}
\label{sec:design}

This section presents the design choices behind the proposed communication path. Starting from the system view above, we specify how dispatch and combine are organized over globally pooled \ac{HBM}, lightweight communication state, and a read-favored execution path on Ascend. The proposed design follows three goals: \textbf{G1}, eliminate or strongly reduce large intermediate relay and reorder buffers on the dispatch/combine path; \textbf{G2}, replace tensor-heavy control with lightweight communication state such as counts, offsets, exchanged addresses, and synchronization symbols; and \textbf{G3}, align the execution path with practical Ascend behavior, especially the benefit of avoiding repeated write-heavy staging in latency-sensitive decode.

\subsection{System Scope}

The proposed implementation is an \ac{MoE}-oriented communication path on Ascend, with primary focus on dispatch and combine. It sits between upper-layer serving frameworks such as vLLM, SGLang, and MindIE, and the lower communication/runtime layer. The implementation relies on globally pooled \ac{HBM} exposed through AclShmem~\cite{aclshmem_repo} or Memfabric~\cite{memfabric_hybrid_repo}, together with Ascend Runtime for synchronization and device execution support. Figure~\ref{fig:zerobufferep-overview} shows this placement.

The stack includes \ac{MoE}-specific dispatch/combine kernels for both \textit{normal} and \textit{low-latency} paths, a small set of supporting collective kernels, and runtime support for memory allocation, buffer registration, barriers, loop counters, core splitting, and related coordination utilities. This report focuses on dispatch and combine because they introduce the main communication difficulty of \ac{MoE} inference: irregular token routing, dynamic expert imbalance, layout conversion, and decode-time latency sensitivity.

\subsection{Pooled-HBM Memory Foundation}

The proposed implementation is built on an existing lower-level capability on Ascend: globally pooled \ac{HBM} exposed through symmetric-memory allocation and shmem-style remote access. This capability is not the primary contribution of this work, but it is essential to the implementation because it provides a stable communication-domain-wide address view for destination expert windows, remote expert windows, and compact communication state.

At bootstrap time, the communication domain establishes a world-visible symmetric device virtual address space. Each rank derives its local memory slice through deterministic offsetting within this world-visible space, rather than through ad hoc payload-specific pointer exchange. This gives the implementation a stable and structured global memory view over which communication can be organized.

Each rank-local slice is further partitioned into two functional regions: a prefix region reserved for communicator-visible state and a suffix region managed as an allocator-controlled heap. As a result, communication-visible state and runtime payload memory are built on top of the same pooled \ac{HBM} basis rather than being managed as disconnected subsystems. This organization allows communication control, tensor storage, and runtime coordination to share one consistent memory foundation.

The role of this memory foundation is to remove the assumption that communication must always be expressed as movement between isolated local sender and receiver buffers. Once a globally visible memory view already exists, communication can instead be organized as structured remote access and placement over globally visible memory regions. This directly supports \textbf{G1}: reducing the need for large explicit relay and restore buffers that exist mainly because of a local-buffer-centric formulation.

\begin{figure}[t]
    \centering
    \includegraphics[width=0.8\linewidth]{"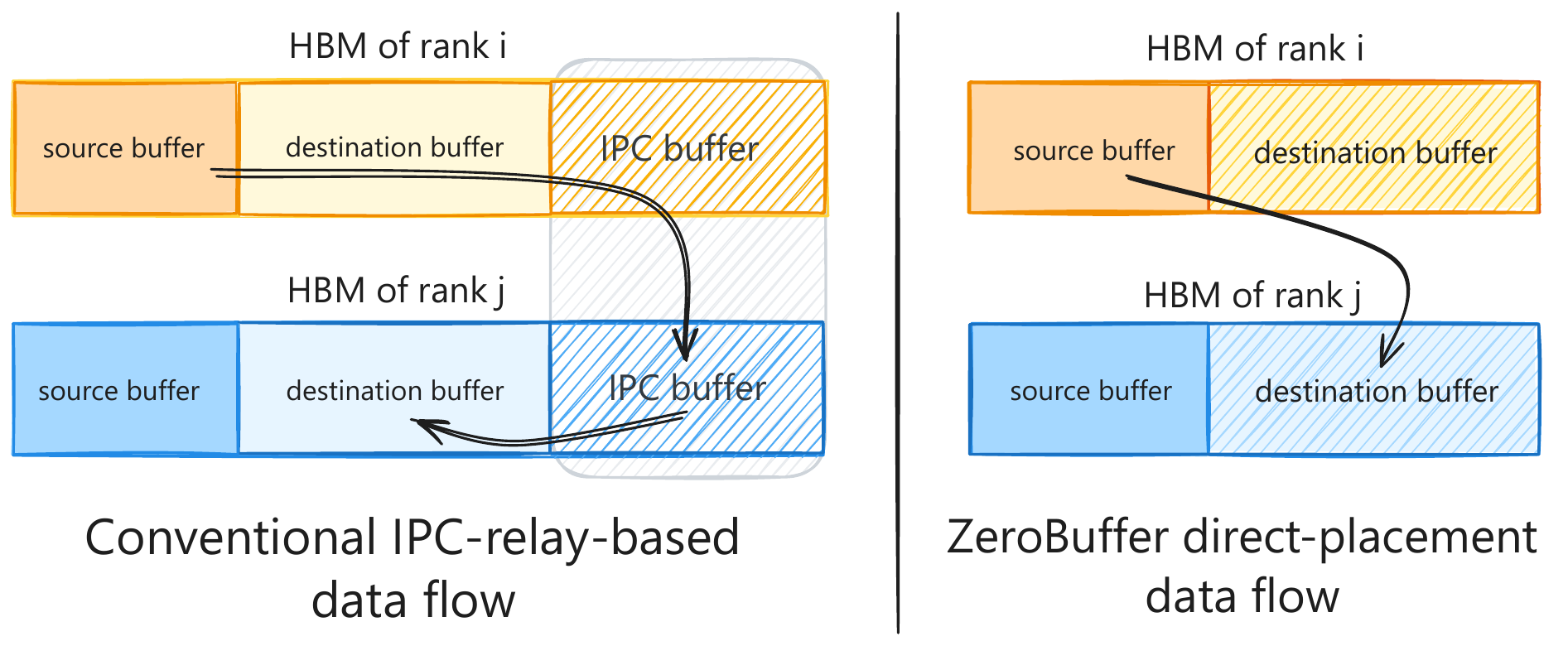"}
    \caption{Comparison between a conventional IPC-relay-based data flow and the proposed direct-placement data flow. In the conventional design, routed payloads are first written into an intermediate IPC relay buffer and later restored into the destination-visible working buffer. In the proposed path, payloads are placed directly into the destination-visible working buffer, removing the IPC relay buffer from the main payload path.}
    \label{fig:zbep-dataflow}
\end{figure}

\subsection{Relay-Buffer-Free Execution Model and Communication State}

The pooled \ac{HBM} foundation leads to a relay-buffer-free execution model. Its basic principle is to keep control state compact and let payload movement happen through direct access to communication-visible memory rather than through large intermediate tensors.

Conventional \ac{IPC} relay buffers are attractive because they provide a communication-visible exchange region, a temporary layout-conversion target, and a synchronization anchor for producer-consumer coordination. However, this also turns the relay buffer into the center of a pack-relay-reorder-restore pipeline. As a result, the system pays not only for communication itself, but also for repeated temporary writes, extra reads, and additional synchronization around the relay stage. The proposed design weakens this dependency by using globally visible memory together with lightweight communication state to move data closer to its final destination earlier in the pipeline. Figure~\ref{fig:zbep-dataflow} illustrates this difference.

To achieve \textbf{G2}, the proposed design separates payload movement from communication control. Payloads are placed in communication-visible memory, while execution is guided by compact communication state such as counts, offsets, exchanged addresses, synchronization symbols, and status flags. In our implementation, such state includes lightweight synchronization and execution-control mechanisms such as wait symbols, vector barriers, and core barriers, together with compact communication buffers for address exchange, count propagation, and offset/status tracking. These states are sufficient to coordinate routed token movement without requiring full intermediate tensor staging.

The normal and low-latency paths instantiate this principle differently. The normal path tolerates more explicit layout-related state, which is acceptable in throughput-oriented workloads such as prefill. The low-latency path compresses communication control into compact state such as counts, offsets, exchanged addresses, expert-window placement information, and lightweight status flags, so that tokens or expert outputs move closer to their final consumer-visible layout earlier in the pipeline.

``Relay-buffer-free'' does \textbf{not} mean zero state. Synchronization state, compact communication buffers, and local temporary storage still exist. The term refers specifically to eliminating or strongly reducing the large explicit relay, staging, and reorder buffers that would otherwise sit on the critical path of \ac{MoE} dispatch and combine.

\subsection{Read-favored Dispatch and Combine}

To achieve \textbf{G3}, the proposed implementation further refines the execution path according to practical hardware behavior on Ascend. A key observation from implementation and deployment is that read-favored execution is often preferable to repeatedly writing temporary staging layouts before final consumption. This is particularly relevant in low-latency \ac{MoE} communication.

The proposed implementation reflects this observation differently in dispatch and combine. For dispatch, the low-latency path is designed to place tokens directly into the destination expert window once the necessary address and count information is available. Instead of first constructing a separate large staging tensor and then performing another placement step, the destination expert buffer itself becomes the semantic target of communication. This shortens the data transition path and directly supports \textbf{G1} and \textbf{G3}.

For combine, the proposed implementation adopts a more explicitly read-favored path. Rather than requiring producers to push outputs through another explicit restore pipeline, the consumer side uses communication state to locate remote expert outputs, reads them from producer-visible expert windows, and performs local weighted reduction and output assembly. This mapping is natural for combine because the final output order is owned by the consumer side. In addition, once remote address information has been exchanged and cached, repeated address handshakes can be shortened through a pre-exchanged fast path. This further reduces the critical path of latency-sensitive combine.

These choices express a broader design principle: communication should be aligned with how the hardware is most naturally used, while reducing the number of explicit intermediate states around dispatch and combine. In this sense, read-favored dispatch and combine are direct realizations of \textbf{G3}.
\section{Prefill and Decode Schedules}
\label{sec:prefill-decode}

The proposed implementation follows the practical split between throughput-oriented \textbf{Prefill} and latency-critical \textbf{Decode} used in DeepEP-style \ac{MoE} communication~\cite{deepep}. The split itself is not new; our focus is to apply the relay-buffer-free communication model to both schedules. Both schedules implement dispatch through direct placement into destination expert windows and combine through direct reading from remote expert windows, but they differ in how much planning and coordination state they retain.

Prefill usually processes a larger token set per invocation, so richer routing analysis, count propagation, and placement planning can be amortized over more token branches. Reducing relay and reordering buffers is also important for prefill because long-context and large-batch execution increase transient \ac{HBM} pressure. Decode, in contrast, usually operates on smaller token sets but lies directly on the serving critical path. It therefore benefits from a narrower control path, fewer repeated coordination steps, and a more compact low-latency communication procedure.

\begin{figure}[!tbp]
    \centering
    \begin{subfigure}[t]{0.47\linewidth}
        \centering
        \includegraphics[width=\linewidth]{"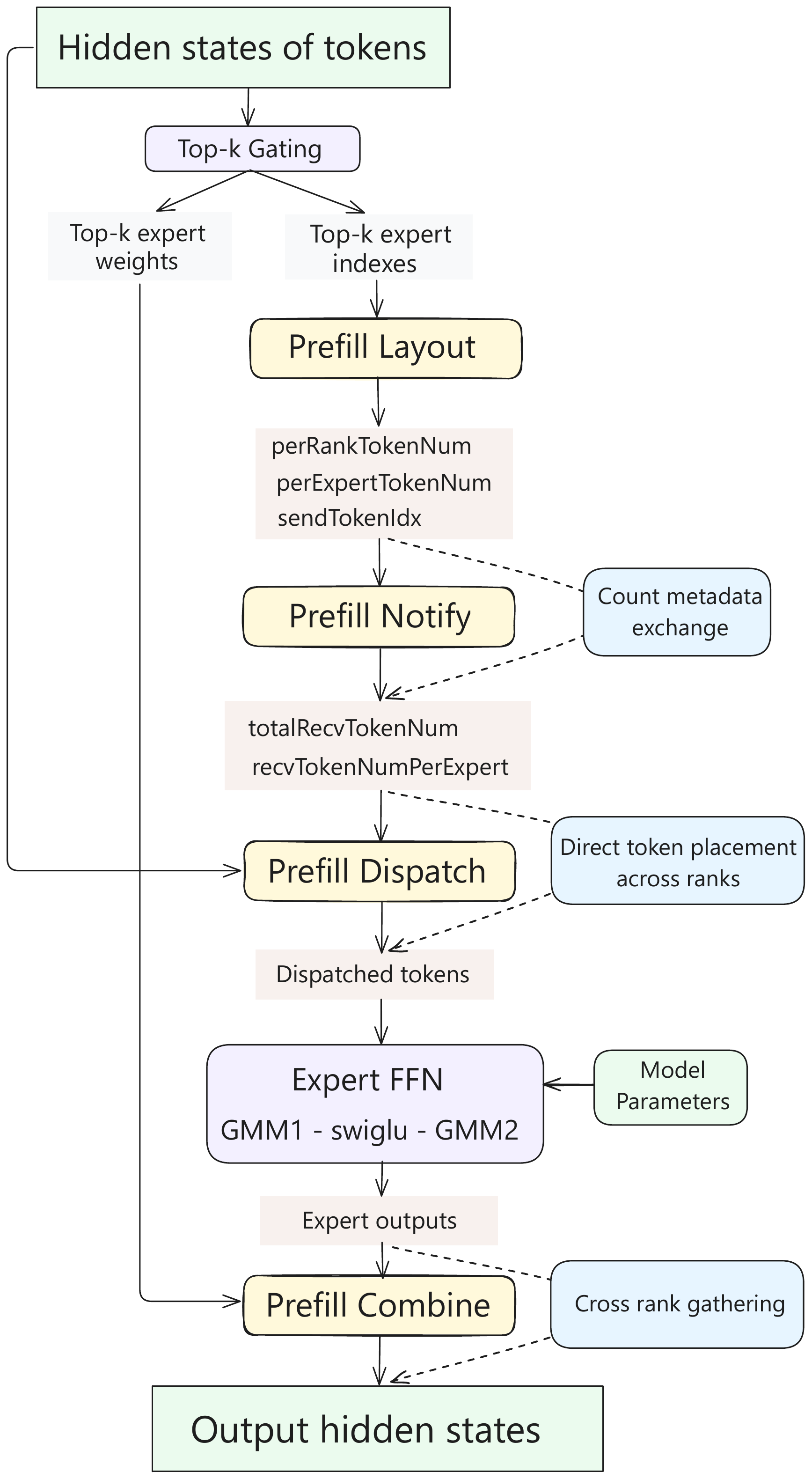"}
        \caption{Workflow of the prefill schedule in the proposed implementation. Starting from the hidden states of tokens, top-$k$ gating produces routing indexes and weights. \textbf{Prefill Layout} computes routing metadata, and \textbf{Prefill Notify} derives count-based placement information through count metadata exchange. \textbf{Prefill Dispatch} then performs direct token placement across ranks. After expert computation in the expert \acp{FFN}, \textbf{Prefill Combine} aggregates expert outputs and restores output hidden states.}
        \label{fig:prefill-workflow}
    \end{subfigure}
    \hfill
    \begin{subfigure}[t]{0.47\linewidth}
        \centering
        \includegraphics[width=\linewidth]{"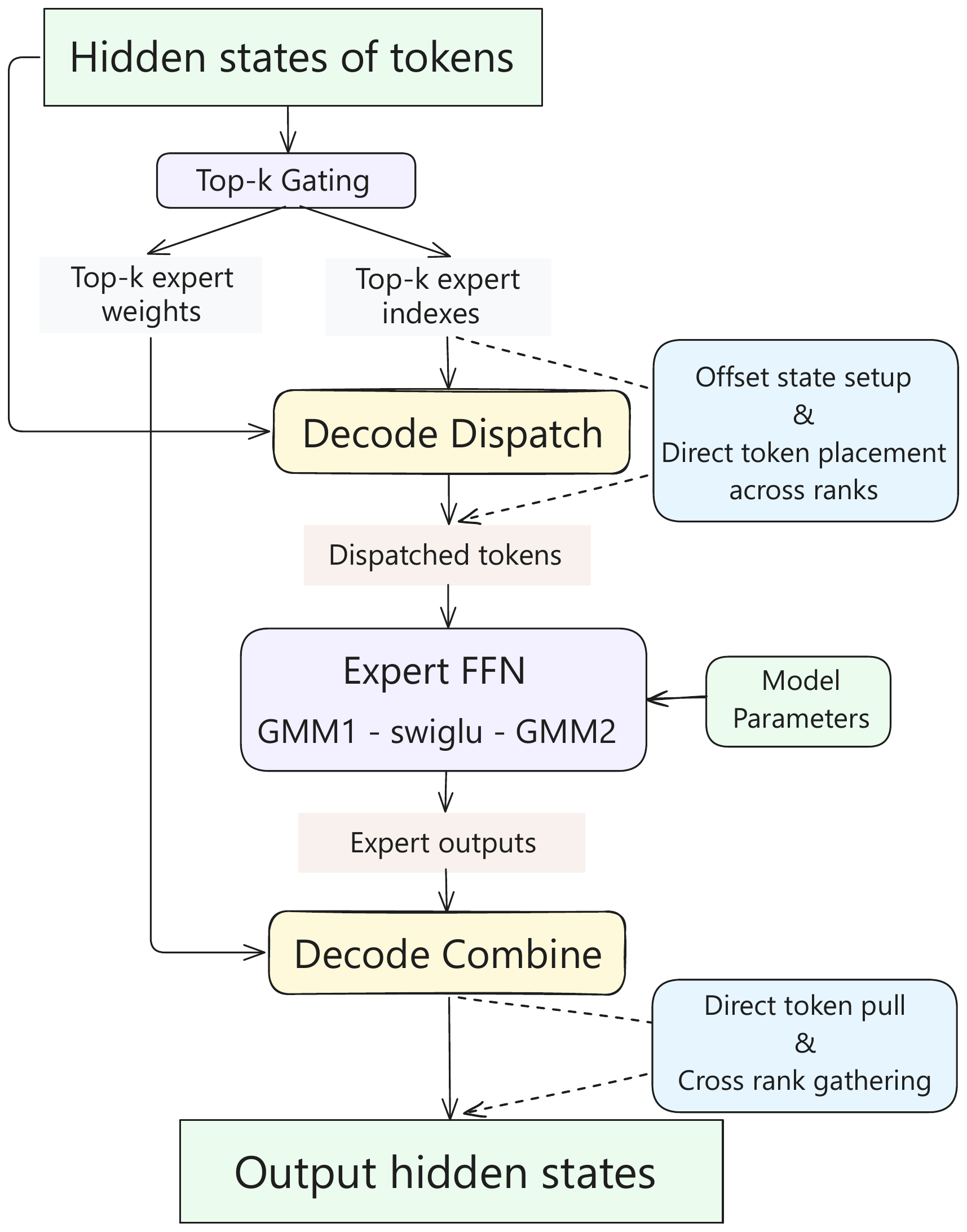"}
        \caption{Workflow of the decode schedule in the proposed implementation. Starting from the hidden states of tokens, top-$k$ gating produces routing indexes and weights. \textbf{Decode Dispatch} uses lightweight offset-related state and performs direct token placement across ranks into destination expert windows. After expert computation in the expert \acp{FFN}, \textbf{Decode Combine} directly reads remote expert outputs, aggregates them across ranks, and restores output hidden states.}
        \label{fig:decode-workflow}
    \end{subfigure}
    \caption{Workflows of the proposed implementation in the two main phases of \ac{MoE} inference. The prefill schedule preserves richer planning state through \textbf{Prefill Layout} and \textbf{Prefill Notify}, while the decode schedule compresses more coordination into \textbf{Decode Dispatch} and \textbf{Decode Combine}. In both cases, dispatch performs direct token placement into destination expert windows, and combine restores output hidden states through direct reading of remote expert outputs.}
    \label{fig:zbep-workflows}
\end{figure}

\subsection{Prefill Schedule}

The prefill schedule is designed for large-batch \ac{MoE} inference and other high-throughput scenarios. Its execution procedure follows the explicit four-stage sequence
\[
\texttt{Prefill\ Layout}
\rightarrow
\texttt{Prefill\ Notify}
\rightarrow
\texttt{Prefill\ Dispatch}
\rightarrow
\texttt{Prefill\ Combine}.
\]
Figure~\ref{fig:prefill-workflow} illustrates this procedure.

\textbf{Prefill Layout} performs local routing analysis and computes routing metadata such as per-rank routed-branch counts, per-expert routed-branch counts, and token-local ordering indexes. \textbf{Prefill Notify} then propagates count metadata across ranks and derives placement state such as receive counts and put offsets. These two stages exchange metadata and coordination state rather than routed token payloads, so they prepare the placement plan without constructing a packed payload relay tensor.

With these planning stages in place, \textbf{Prefill Dispatch} performs direct placement. Once destination tensor addresses are available, each token is written directly into the destination expert window. The destination row is determined by the destination rank together with a placement offset of the form
\[
\texttt{putOffset} + \texttt{sendTokenIdx}.
\]
Thus, payloads are placed directly into the semantic expert tensor rather than into an intermediate relay region.

\textbf{Prefill Combine} performs direct reading. For each token and each top-$k$ branch, the kernel computes the remote row from count-derived base offsets together with the token-local index,
\[
\texttt{epRecvCount} + \texttt{sendTokenIdx},
\]
then reads that row directly from the remote expert window and performs local weighted reduction. The prefill schedule therefore realizes relay-buffer-free communication through direct placement and direct reading, while retaining richer routing, counting, and offset-construction state that can be amortized over larger token counts.

\subsection{Decode Schedule}

The decode schedule follows the same relay-buffer-free communication model, but applies it with a more compact control procedure. Compared with prefill, decode removes the explicit \texttt{Layout}$\rightarrow$\texttt{Notify} sequence and folds the required count, offset, address, and synchronization state into the dispatch/combine procedure. This reduces the amount of host-visible routing state and repeated coordination on the latency-critical path.

In \textbf{Decode Dispatch}, once the destination rank, expert-level cumulative offset, and token-local offset are known, the kernel computes the final destination address and writes the token directly into the destination expert window. In implementation terms, this placement follows the same direct expert-window idea as prefill, but with a tighter addressing rule based on
\[
\texttt{dstExpertOffset} + \texttt{curExpertCnt}.
\]
The destination expert buffer itself remains the semantic target of communication; the difference is that decode uses a narrower control path and a shorter communication process.

\textbf{Decode Combine} is the most compact read-favored realization of the proposed communication model. Instead of restoring payloads from a generic relay region, the consumer side locates the required remote expert outputs, reads them directly from producer-visible expert windows, and performs local weighted reduction. Once the relevant remote-window address and offset state are available, each consumer token uses its expert assignment, expansion index, send-count-derived prefix offset, and scaling information to compute the remote row. The address is formed as
\[
\texttt{remoteBase} + \texttt{remoteOffset},
\]
where the base comes from expert-level token-count state and the offset comes from token-local expansion state.

This combine procedure is read-favored because the dominant data movement is initiated by the consumer, which reads exactly the expert outputs it needs and assembles the final output locally. Optional cached-address paths can further shorten repeated coordination when remote-window addresses are already available; the concrete address-exchange and compatibility details are implementation choices discussed in Section~\ref{sec:implementation}.

\subsection{Schedule Comparison}

Although Prefill and Decode use different schedules, they share the same relay-buffer-free communication model. Both rely on the same pooled-\ac{HBM} memory foundation and avoid using a generic payload relay buffer as the primary carrier of routed token movement. In both cases, dispatch performs direct placement into expert windows, while combine performs direct reading from expert windows followed by local reduction.

The main difference is the amount of planning and coordination state retained by each schedule. The prefill schedule follows the DeepEP-style throughput path and keeps richer layout, count, and offset state, because the planning overhead can be amortized over larger token sets and can help reduce transient memory pressure in long-context or large-batch execution. The decode schedule follows the DeepEP-style low-latency path and narrows the exposed control state to the count, offset, address, and synchronization information needed on the serving critical path. Thus, the proposed design should be understood as a unified communication model for \ac{MoE} dispatch and combine, instantiated as two kernel schedules tailored to the two main phases of \ac{MoE} inference.
\section{Implementation}
\label{sec:implementation}

The proposed design is implemented as a set of \ac{MoE} communication kernels on top of globally pooled \ac{HBM} exposed through symmetric-memory allocation and shmem-style remote access. At the implementation level, the key issue is not only where token payloads are moved, but also how routing results are transformed into placement offsets, expert-window layouts, and final token-aligned outputs. For this reason, the implementation is best understood through the transformation of tensors and lightweight communication states across the prefill and decode schedules.

Although the two schedules differ in the amount of explicit planning state they preserve, both implement the same payload movement rule: dispatch writes token rows directly into destination expert windows, and combine restores token-aligned outputs by directly reading expert outputs from remote expert windows.

\subsection{Notation and Intermediate States}

Let \(T\) denote the number of input tokens, \(H\) the hidden size, \(k\) the top-\(k\) routing degree, \(E\) the total number of experts, and \(R\) the number of ranks. The input hidden states are denoted by \(X \in \mathbb{R}^{T \times H}\), the routing indexes by \(K \in \{0,\dots,E-1\}^{T \times k}\), and the routing weights by \(W \in \mathbb{R}^{T \times k}\). For the \(j\)-th routed branch of token \(t\), let \(K_{t,j}\) be its destination expert, and let
\[
r_{t,j} = \Big\lfloor \frac{K_{t,j}}{E_r} \Big\rfloor
\]
be the destination rank, where \(E_r\) is the number of experts per rank.

At the implementation level, the most important intermediate states are two kinds of offsets.

First, for each routed branch \((t,j)\), the implementation computes a \emph{small offset} \(s_{t,j}\), which records the token-local position of that branch within the receive stream of its destination expert. In the current codebase, this quantity is represented by \texttt{sendTokenIdx} in the prefill schedule and by \texttt{expandIdx} in the decode schedule.

Second, the implementation computes a \emph{large offset} \(o_{e,r}\), which records where the block of routed branches sent from source rank \(r\) to expert \(e\) begins in the destination expert window. In the current implementation, this role is played by \texttt{putOffset} in prefill and by \texttt{ep\_recv\_count} in decode.

Using these two offsets, both schedules locate a token row in an expert window through the same general rule:
\[
\text{expert-window row} = o_{e,r} + s_{t,j}.
\]
This rule is the key implementation mechanism behind direct placement in dispatch and direct reading in combine.

Table~\ref{tab:zbep-tensors} summarizes representative tensors and lightweight states used by the proposed prefill and decode implementations. The exact allocation and layout are implementation-dependent, but the table captures the main state categories used to construct expert-window rows and restore token-aligned outputs.

\begin{table}[t]
\centering
\caption{Representative tensors and lightweight states used by the proposed prefill and decode implementations.}
\label{tab:zbep-tensors}
\begin{tabular}{llll}
\toprule
Symbol & Code name & Meaning & Shape \\
\midrule
\(X\) & \texttt{x} & input hidden states & \(T \times H\) \\
\(K\) & \texttt{topkIdx} & top-\(k\) routing indexes & \(T \times k\) \\
\(W\) & \texttt{topkWeights} & top-\(k\) routing weights & \(T \times k\) \\
\(c^{\mathrm{rank}}\) & \texttt{perRankTokenNum} & routed branches sent to each rank & \(R\) \\
\(c^{\mathrm{exp}}\) & \texttt{perExpertTokenNum} & routed branches assigned to each expert & \(E\) \\
\(s\) & \texttt{sendTokenIdx} & local offset in expert stream (prefill) & \(T \times k\) \\
\(M\) & \texttt{recvData} & gathered count matrix & \(R \times E\) \\
\(o\) & \texttt{putOffset} & large offset table (prefill) & implementation-dependent \\
\(n^{\mathrm{recv}}\) & \texttt{totalRecvTokenNum} & total received routed branches & scalar \\
\(n^{\mathrm{exp}}\) & \texttt{recvTokenNumPerExpert} & received routed branches per local expert & local experts \\
\(X^{\mathrm{exp}}\) & \texttt{expandXOut} & dispatched expert-window tensor & implementation-dependent \\
\(\delta\) & \texttt{expandIdx} & local offset in expert stream (decode) & implementation-dependent \\
\(o^{\mathrm{dec}}\) & \texttt{ep\_recv\_count} & large offset table (decode) & implementation-dependent \\
\(Y\) & \texttt{combined\_x} & output hidden states & \(T \times H\) \\
\bottomrule
\end{tabular}
\end{table}

\subsection{Implementation of the Prefill Schedule}

The implementation of the prefill schedule is organized as a sequence of tensor transformations shown in Figure~\ref{fig:prefill-workflow}. Starting from the input hidden states \(X\), routing indexes \(K\), and routing weights \(W\), the schedule first derives routing metadata, then converts count metadata into placement state, next writes token rows into destination expert windows, and finally restores token-aligned outputs through direct remote reading and weighted reduction.

\begin{algorithm}[t]
\caption{Prefill implementation of the proposed communication path}
\label{alg:prefill-zbep}
\begin{algorithmic}[1]
\Require hidden states \(X\), routing indexes \(K\), routing weights \(W\)
\Ensure token-aligned outputs \(Y\)

\State \textbf{Layout:} compute per-rank routed-branch counts \(c^{\mathrm{rank}}\), per-expert routed-branch counts \(c^{\mathrm{exp}}\), and token-local offsets \(s\)
\State \textbf{Notify:} exchange count metadata and construct expert-window base offsets \(o\)
\State allocate destination expert-window tensor \(X^{\mathrm{exp}}\)

\ForAll{routed branches \((t,j)\)}
    \State \(e \leftarrow K_{t,j}\), \(r_{\mathrm{src}} \leftarrow\) source rank of token \(t\)
    \State \(p_{t,j} \leftarrow o_{e,r_{\mathrm{src}}} + s_{t,j}\)
    \State directly write \(X_t\) into row \(p_{t,j}\) of the destination expert window
\EndFor

\State run expert \acp{FFN} on \(X^{\mathrm{exp}}\) to produce \(Y^{\mathrm{exp}}\)
\State initialize \(Y \leftarrow 0\)

\ForAll{routed branches \((t,j)\)}
    \State \(e \leftarrow K_{t,j}\), \(r_{\mathrm{dst}} \leftarrow \lfloor e/E_r \rfloor\)
    \State \(q_{t,j} \leftarrow o_{e,r_{\mathrm{dst}}} + s_{t,j}\)
    \State directly read \(Y^{\mathrm{exp}}_{q_{t,j}}\) from the remote expert window
    \State \(Y_t \leftarrow Y_t + W_{t,j} \cdot Y^{\mathrm{exp}}_{q_{t,j}}\)
\EndFor

\State \Return \(Y\)
\end{algorithmic}
\end{algorithm}

\paragraph{Prefill Layout: from routing indexes to routing metadata.}
The input of \textbf{Prefill Layout} is the routing tensor \(K\). This stage does not move payload rows from \(X\). Instead, it converts routing results into explicit metadata required by later communication stages.

For each token \(t\) and each routed branch \(j\), the kernel reads the destination expert \(K_{t,j}\), derives its destination rank \(r_{t,j}\), and updates two counting tensors:
\[
c^{\mathrm{rank}}_r
=
\sum_{t,j} \mathbf{1}[r_{t,j}=r],
\qquad
c^{\mathrm{exp}}_e
=
\sum_{t,j} \mathbf{1}[K_{t,j}=e].
\]
In the current implementation, these correspond to \texttt{perRankTokenNum} and \texttt{perExpertTokenNum}, respectively. Both are routed-branch counts: \texttt{perRankTokenNum} aggregates routed branches by destination rank, while \texttt{perExpertTokenNum} aggregates routed branches by destination expert.

At the same time, the layout kernel computes the token-local ordering of each routed branch within its destination expert stream. Denoting this local order by \(s_{t,j}\), we have
\[
s_{t,j}
=
\#\{(t',j') \text{ before } (t,j) \mid K_{t',j'} = K_{t,j}\}.
\]
This quantity is represented in the current code by \texttt{sendTokenIdx}. Operationally, \texttt{sendTokenIdx} is obtained through per-block counting plus prefix accumulation over earlier blocks, rather than through any payload reordering.

Therefore, after \textbf{Prefill Layout}, the implementation has produced routing metadata
\[
\bigl(
c^{\mathrm{rank}},
c^{\mathrm{exp}},
s
\bigr),
\]
without constructing any packed relay tensor for the payload itself.

\paragraph{Prefill Notify: from local counts to global placement state.}
The input of \textbf{Prefill Notify} is the count metadata generated by \textbf{Prefill Layout}, in particular the per-expert count tensor \(c^{\mathrm{exp}}\). This stage exchanges count metadata across ranks and transforms them into placement offsets and scheduling information.

Concretely, the implementation first gathers all per-rank expert counts into a matrix
\[
M \in \mathbb{N}^{R \times E},
\qquad
M_{r,e} \text{ stands for the number of routed branches sent from rank } r \text{ to expert } e,
\]
which corresponds to \texttt{recvData} in the current implementation.

It then constructs the large-offset tensor \(o_{e,r}\), which records the starting row of the block sent from source rank \(r\) to expert \(e\) in the destination expert window. In code, this tensor is represented by \texttt{putOffset}. The notify stage also derives:
\begin{itemize}
    \item the total number of received routed branches at the current rank, represented by \texttt{totalRecvTokenNum},
    \item the number of received routed branches for each local expert, represented by \texttt{recvTokenNumPerExpert},
    \item an auxiliary scheduling tensor for helper processing, represented by \texttt{balanceMatrix}.
\end{itemize}

Thus, \textbf{Prefill Notify} transforms local count metadata into a global placement rule. If \textbf{Prefill Layout} answers ``which expert does a routed branch belong to,'' then \textbf{Prefill Notify} answers ``where does that branch begin in the destination expert window.''

\paragraph{Prefill Dispatch: from input hidden states to destination expert windows.}
The input of \textbf{Prefill Dispatch} consists of the payload tensor \(X\), the routing indexes \(K\), the small offsets \(s_{t,j}\), and the large offsets \(o_{e,r}\). The output is the expert-window tensor \(X^{\mathrm{exp}}\), represented in the current implementation by \texttt{expandXOut}.

For each routed branch \((t,j)\), dispatch computes the destination row as
\[
p_{t,j}
=
o_{K_{t,j},\,r_{\mathrm{src}}}
+
s_{t,j},
\]
where \(r_{\mathrm{src}}\) is the source rank of token \(t\). The hidden-state row \(X_t\) is then written directly into row \(p_{t,j}\) of the destination expert window.

The important implementation point is that \(X^{\mathrm{exp}}\) is not assembled by first writing routed rows into an intermediate relay tensor and then restoring them into expert order. Instead, the destination expert window itself is the immediate target of dispatch. If row-wise quantization is enabled, the corresponding scale values are written into a parallel scale tensor in the same row order.

As a result, after \textbf{Prefill Dispatch}, the payload transformation is:
\[
X
\longrightarrow
X^{\mathrm{exp}},
\]
where \(X^{\mathrm{exp}}\) is already organized in expert-window order.

\paragraph{Expert computation.}
The expert \acp{FFN} directly consume \(X^{\mathrm{exp}}\) and produce expert outputs \(Y^{\mathrm{exp}}\). No additional relay-style reordering is needed between dispatch and expert computation.

\paragraph{Prefill Combine: from expert outputs back to token-aligned outputs.}
The input of \textbf{Prefill Combine} consists of the expert-window output tensor \(Y^{\mathrm{exp}}\), the routing weights \(W\), the routing indexes \(K\), and the two offset states reused from dispatch. The output is the token-aligned hidden-state tensor \(Y \in \mathbb{R}^{T \times H}\).

For each token \(t\) and routed branch \(j\), combine locates the corresponding expert-output row through
\[
q_{t,j}
=
o_{K_{t,j},\,r_{\mathrm{dst}}}
+
s_{t,j},
\]
which in the current implementation corresponds to the sum of the expert-level base offset and \texttt{sendTokenIdx}. The combine kernel then reads row \(q_{t,j}\) directly from the remote expert window, multiplies it by the routing weight \(W_{t,j}\), and accumulates over all selected branches:
\[
Y_t
=
\sum_{j=1}^{k}
W_{t,j}\,
Y^{\mathrm{exp}}_{q_{t,j}}.
\]

Hence, the full tensor transformation chain of the prefill schedule is:
\[
(X, K, W)
\rightarrow
(c^{\mathrm{rank}}, c^{\mathrm{exp}}, s)
\rightarrow
(M, o)
\rightarrow
X^{\mathrm{exp}}
\rightarrow
Y^{\mathrm{exp}}
\rightarrow
Y.
\]

In this sense, the prefill schedule preserves richer planning state, but payload movement itself already follows direct placement in dispatch and direct reading in combine.

\subsection{Implementation of the Decode Schedule}

Compared with the prefill schedule, the decode schedule removes the explicit \textbf{Prefill Layout} and \textbf{Prefill Notify} stages and compresses more logic into the two kernel stages \textbf{Decode Dispatch} and \textbf{Decode Combine}. Nevertheless, the same two-level offset rule is still preserved.

\begin{algorithm}[t]
\caption{Decode implementation of the proposed communication path}
\label{alg:decode-zbep}
\begin{algorithmic}[1]
\Require hidden states \(X\), routing indexes \(K\), routing weights \(W\)
\Ensure token-aligned outputs \(Y\)

\State compute compact count state, decode base offsets \(o^{\mathrm{dec}}\), and token-local offsets \(\delta\)
\State prepare required destination and remote expert-window addresses
\State allocate decode expert-window tensor \(X^{\mathrm{exp}}_{\mathrm{dec}}\)

\ForAll{routed branches \((t,j)\)}
    \State \(e \leftarrow K_{t,j}\), \(r_{\mathrm{src}} \leftarrow\) source rank of token \(t\)
    \State \(p^{\mathrm{dec}}_{t,j} \leftarrow o^{\mathrm{dec}}_{e,r_{\mathrm{src}}} + \delta_{t,j}\)
    \State directly write \(X_t\) into row \(p^{\mathrm{dec}}_{t,j}\) of the destination expert window
\EndFor

\State run expert \acp{FFN} on \(X^{\mathrm{exp}}_{\mathrm{dec}}\) to produce \(Y^{\mathrm{exp}}_{\mathrm{dec}}\)
\State initialize \(Y \leftarrow 0\)

\ForAll{routed branches \((t,j)\)}
    \State \(e \leftarrow K_{t,j}\), \(r_{\mathrm{dst}} \leftarrow \lfloor e/E_r \rfloor\)
    \State \(q^{\mathrm{dec}}_{t,j} \leftarrow b_{e,r_{\mathrm{dst}}} + \delta_{t,j}\)
    \State directly read \(Y^{\mathrm{exp}}_{\mathrm{dec},\,q^{\mathrm{dec}}_{t,j}}\) from the remote expert window
    \State \(Y_t \leftarrow Y_t + W_{t,j} \cdot Y^{\mathrm{exp}}_{\mathrm{dec},\,q^{\mathrm{dec}}_{t,j}}\)
\EndFor

\State \Return \(Y\)
\end{algorithmic}
\end{algorithm}

\paragraph{Decode Dispatch: from routing indexes to compact placement state and expert-window inputs.}
The inputs of \textbf{Decode Dispatch} are again the input hidden states \(X\), the routing indexes \(K\), and the routing weights \(W\). However, instead of constructing a separate richer planning chain, decode derives only the minimum count and offset state required for direct placement.

For each routed branch \((t,j)\), the dispatch procedure first counts how many routed branches are assigned to each expert. These local counts are then combined into a compact global count state, which is used to construct the expert-level large offsets \(o^{\mathrm{dec}}_{e,r}\). In the current implementation, this large-offset tensor is represented by \texttt{ep\_recv\_count}.

At the same time, for each routed branch, decode dispatch computes a token-local small offset \(\delta_{t,j}\), which records the local order of the routed branch within the receive stream of its destination expert. In the current implementation, this is represented by \texttt{expandIdx}.

Using these two quantities, decode dispatch places token \(t\) into the destination expert window at
\[
p^{\mathrm{dec}}_{t,j}
=
o^{\mathrm{dec}}_{K_{t,j},\,r_{\mathrm{src}}}
+
\delta_{t,j}.
\]
Thus, as in prefill, the payload row \(X_t\) is written directly into the destination expert window. The difference is that the large offset \(o^{\mathrm{dec}}\) and the small offset \(\delta\) are generated inside the dispatch procedure itself, rather than by a separate layout-notify chain.

The output of this stage is the dispatched expert-window tensor \(X^{\mathrm{exp}}_{\mathrm{dec}}\), together with the compact offset state \((o^{\mathrm{dec}}, \delta)\).

Compared with the prefill implementation, decode dispatch reduces the host-visible routing tensor chain used by the staged \texttt{Layout}$\rightarrow$\texttt{Notify} sequence. It also resolves only the tensor addresses required for direct placement by default, while selected compatibility or \ac{IPC}-enabled configurations may keep an extra address slot. In addition, decode avoids the more explicit helper-processing schedule used by the throughput-oriented prefill path.

\paragraph{Expert computation.}
The expert \acp{FFN} directly consume \(X^{\mathrm{exp}}_{\mathrm{dec}}\) and produce expert outputs \(Y^{\mathrm{exp}}_{\mathrm{dec}}\).

\paragraph{Decode Combine: from remote expert outputs back to token-aligned outputs.}
The input of \textbf{Decode Combine} consists of \(Y^{\mathrm{exp}}_{\mathrm{dec}}\), the routing weights \(W\), the routing indexes \(K\), and the compact offset state \((o^{\mathrm{dec}}, \delta)\). Decode combine also requires remote expert-window addresses before issuing consumer-side reads. The default standalone wrapper follows a lightweight handshake-based procedure to make the required remote-window addresses available to the kernel. The implementation also supports an optional cached-address mode, in which the consumer directly loads pre-exchanged remote-window addresses and skips repeated per-call address handshakes when the cached state is valid.

For each token \(t\) and routed branch \(j\), decode combine computes the remote row through
\[
q^{\mathrm{dec}}_{t,j}
=
b_{K_{t,j},\,r_{\mathrm{dst}}}
+
\delta_{t,j},
\]
where \(b_{K_{t,j},\,r_{\mathrm{dst}}}\) is the expert-level base offset derived from the count state, and \(\delta_{t,j}\) is the small offset recorded during dispatch. In the current implementation, these correspond to \texttt{remoteBase} and \texttt{remoteOffset}, respectively.

The combine kernel then reads the expert-output row \(Y^{\mathrm{exp}}_{\mathrm{dec},\,q^{\mathrm{dec}}_{t,j}}\) directly from the remote expert window, multiplies it by the routing weight \(W_{t,j}\), and accumulates:
\[
Y_t
=
\sum_{j=1}^{k}
W_{t,j}\,
Y^{\mathrm{exp}}_{\mathrm{dec},\,q^{\mathrm{dec}}_{t,j}}.
\]

Therefore, the tensor transformation chain of decode can be summarized as:
\[
(X, K, W)
\rightarrow
(o^{\mathrm{dec}}, \delta)
\rightarrow
X^{\mathrm{exp}}_{\mathrm{dec}}
\rightarrow
Y^{\mathrm{exp}}_{\mathrm{dec}}
\rightarrow
Y.
\]

Compared with prefill, decode reduces the number of explicit intermediate planning tensors, but the payload transformation rule remains the same: dispatch writes directly into destination expert windows, and combine restores token-aligned outputs by directly reading expert outputs from remote expert windows. Algorithms~\ref{alg:prefill-zbep} and~\ref{alg:decode-zbep} summarize the tensor transformation and communication procedure of the prefill and decode schedules, respectively. The two algorithms make explicit that the main difference between the two schedules lies not in the final payload movement rule, but in how count and offset states are constructed and consumed.

\subsection{Address Exchange and Runtime Coordination}

The proposed implementation removes large payload-carrying relay buffers from the dispatch/combine path, but it still relies on lightweight runtime state for address exchange, synchronization, and progress coordination. In the implementation, this state includes exchanged expert-window addresses, count and offset buffers, wait symbols, vector barriers, core barriers, loop counters, and status flags. These states coordinate when a remote expert window can be written or read, but they do not serve as the primary carrier of routed token payloads.

Selected compatibility configurations may still keep \ac{IPC}-related address slots or auxiliary buffers. These states are used for address compatibility or fallback coordination, rather than as the main payload relay path in the proposed dispatch/combine procedure. This distinction is important: the proposed implementation is not a zero-state implementation, but an implementation that removes or strongly reduces large explicit relay, staging, and reorder buffers from the critical payload path.
\section{Evaluation}
\label{sec:evaluation}

We evaluate the proposed implementation using an early set of kernel-level and serving-level experiments. The goal of this section is not to provide an exhaustive benchmark suite, but to check whether the relay-buffer-free implementation produces consistent evidence across the main execution regimes considered in this report. We evaluate three aspects. First, we measure kernel-level latency for both the prefill and decode communication paths. Second, we examine whether these kernel-level gains are reflected in serving-level metrics, including \ac{TTFT} and \ac{TPOT}. Third, we study whether the improved communication behavior enlarges the feasible scheduling space under practical serving constraints.

\subsection{Experimental Setup}

We compare the proposed implementation against an HCCL-enabled~\cite{hccl} DeepEP baseline under the same execution environment and routing workload. Unless otherwise stated, kernel-level latency is reported in microseconds and end-to-end serving latency is reported in milliseconds.

For decode microbenchmarks, we vary the batch size and study two representative hidden sizes, 4096 and 7168, under both quantized and non-quantized settings. For prefill experiments, we vary the input-token count and separately evaluate the normal-kernel path and a representative low-latency case-study path. For end-to-end serving, we report \ac{TTFT} and \ac{TPOT} on representative DeepSeek serving scenarios~\cite{deepseek2024v3,deepseek_v32}. Unless explicitly stated otherwise, all comparisons use the same routing configuration and runtime environment.

\paragraph{Implementation modes.}
Unless otherwise stated, the experiments use the default prefill-normal and decode-low-latency paths described in Section~\ref{sec:implementation}. Decode combine uses lightweight address handshaking by default. The optional cached-address path is treated as an implementation optimization and is not assumed in the default results unless explicitly reported. Selected compatibility configurations may retain auxiliary \ac{IPC}-related address slots, but these slots are not used as the primary payload relay path.

\subsection{Prefill Normal-Kernel Results}

\begin{figure}[!tbp]
    \centering
    \includegraphics[width=\linewidth]{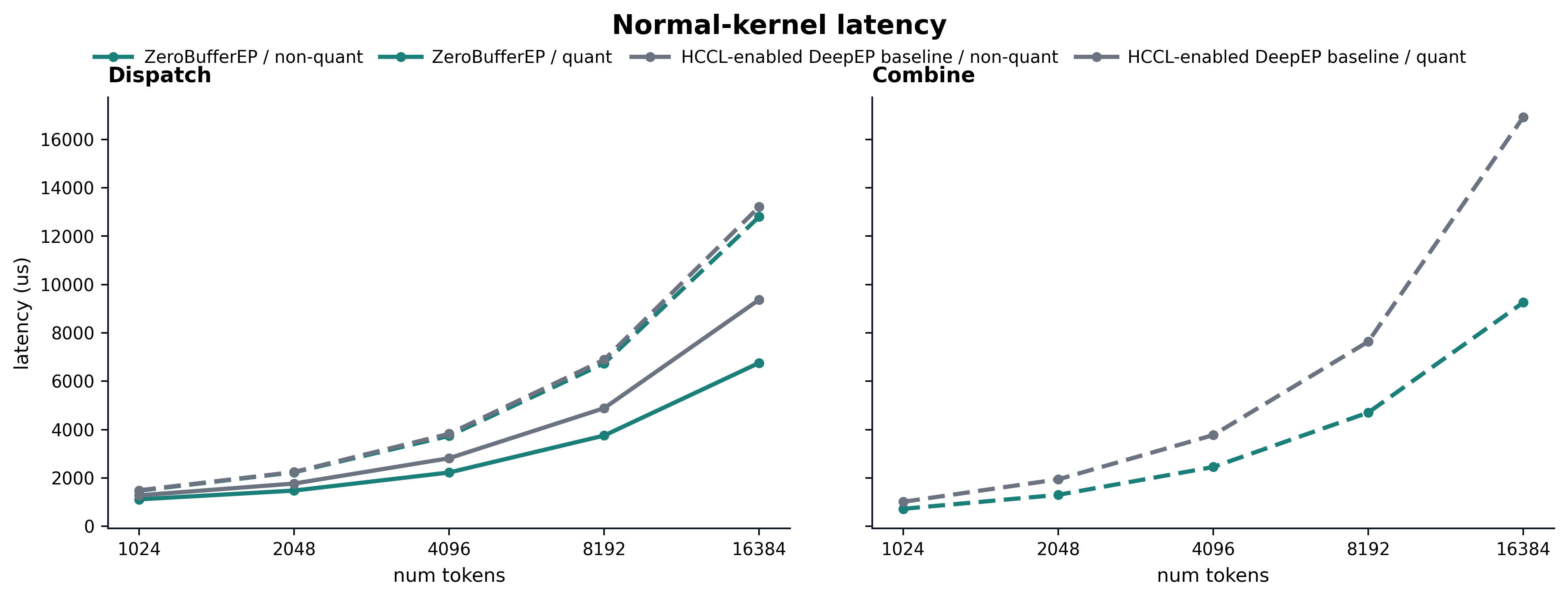}
    \caption{Normal-kernel latency comparison between the proposed implementation and the HCCL-enabled DeepEP baseline for prefill dispatch and combine under different token counts.}
    \label{fig:prefill-normal-kernel}
\end{figure}

We first evaluate the normal-kernel path used in throughput-oriented prefill. Figure~\ref{fig:prefill-normal-kernel} reports dispatch and combine latency as the number of input tokens increases.

\begin{figure}[!tbp]
    \centering
    \includegraphics[width=\linewidth]{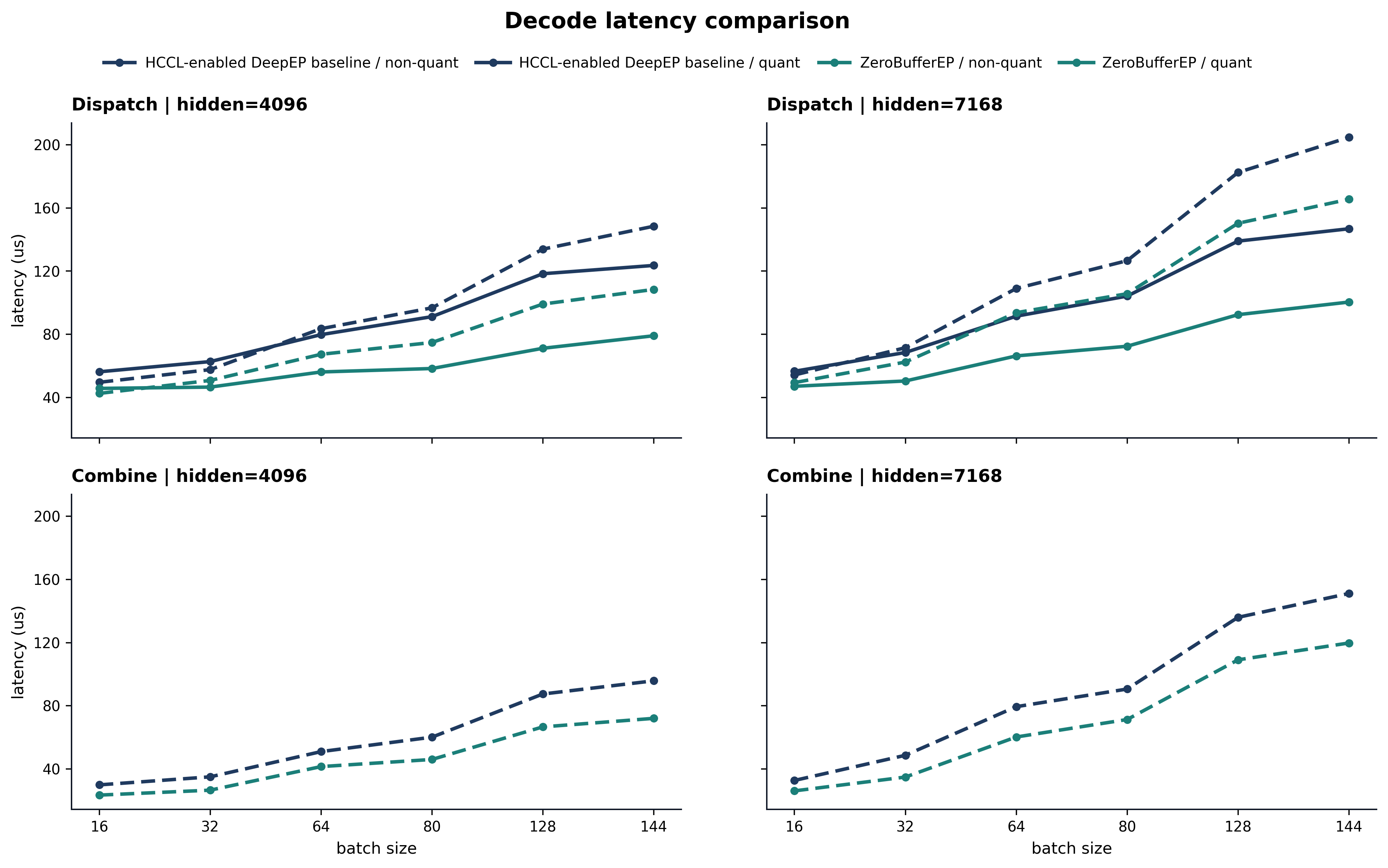}
    \caption{Decode latency comparison between the proposed implementation and the HCCL-enabled DeepEP baseline. The results are reported for dispatch and combine under hidden sizes 4096 and 7168, with both quantized and non-quantized settings.}
    \label{fig:decode-latency-comparison}
\end{figure}

Figure~\ref{fig:prefill-normal-kernel} shows two main trends. First, the proposed implementation reduces both dispatch and combine latency in the non-quantized setting. The gap becomes more visible as the token count grows, which is consistent with the design goal of reducing relay- and reorder-related overhead that would otherwise accumulate in large-token prefill. For example, in dispatch, the baseline increases from about \(1.1\) ms at 1024 tokens to about \(9.4\) ms at 16384 tokens, whereas the relay-buffer-free path increases from about \(1.0\) ms to about \(6.8\) ms. A similar widening gap appears in combine.

Second, under quantization, the gain remains visible but becomes smaller on the dispatch side. A plausible explanation is that quantized execution introduces additional scale-handling work, which reduces the relative share of the relay overhead removed by the proposed path. Even so, the implementation still lowers both dispatch and combine latency in the tested cases, and the advantage becomes clearer at larger token counts.

These results indicate that the proposed path is not limited to decode. The same direct-placement and direct-read principle also benefits the throughput-oriented prefill path, especially when the routed-token volume is large enough for buffer-centric overhead to become visible.

\subsection{Decode Low-Latency Results}

We next evaluate the decode-oriented dispatch and combine kernels. Figure~\ref{fig:decode-latency-comparison} shows the latency trends across batch sizes for hidden sizes 4096 and 7168 under both quantized and non-quantized settings. Table~\ref{tab:decode-summary} summarizes the corresponding reduction over the baseline.

\begin{figure}[!tbp]
    \centering
    \includegraphics[width=\linewidth]{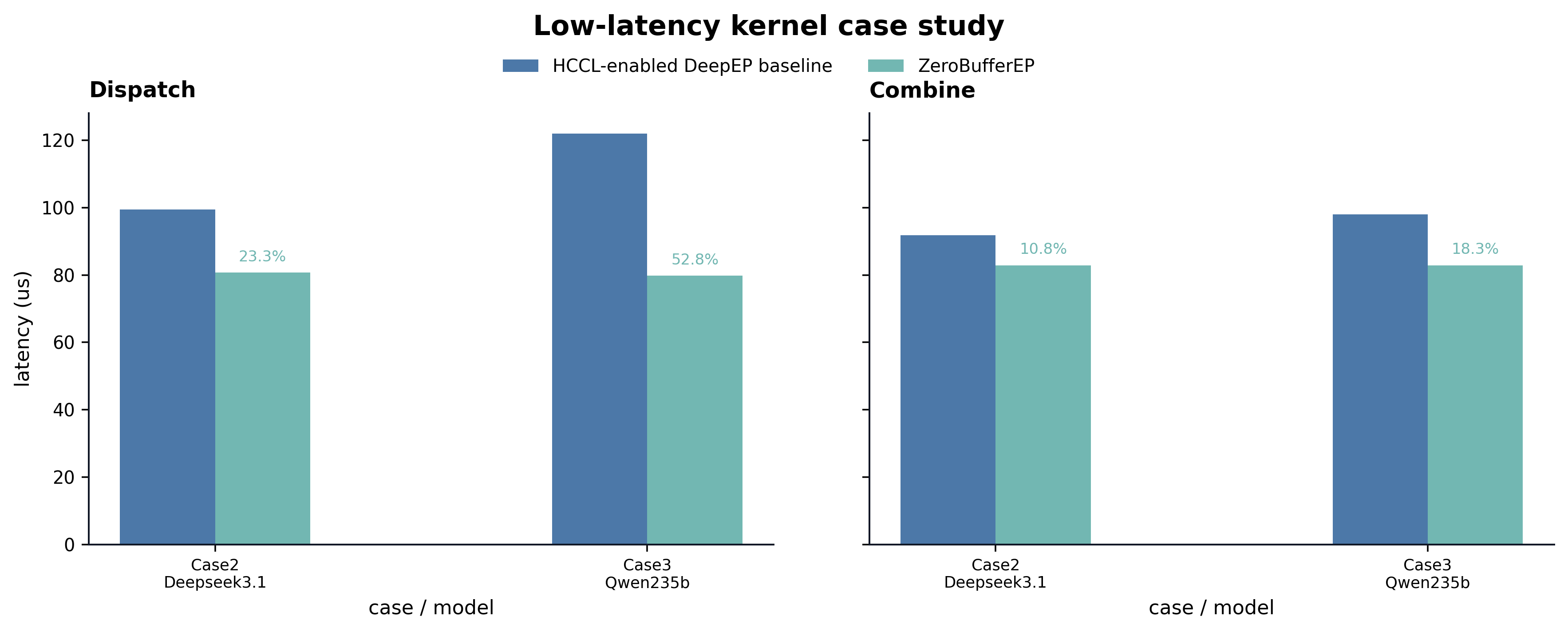}
    \caption{Low-latency kernel case study on representative DeepSeek 3.1 and Qwen-235B settings.}
    \label{fig:lowlatency-case-study}
\end{figure}

Figure~\ref{fig:decode-latency-comparison} and Table~\ref{tab:decode-summary} show that the proposed implementation reduces decode latency for both dispatch and combine across the tested settings. This provides direct evidence that the same payload-movement rule used in prefill remains effective in the latency-sensitive decode regime.

The reduction is especially pronounced for dispatch. In the quantized setting, the average speedup reaches \(31.02\%\) for hidden size 4096 and \(27.72\%\) for hidden size 7168. Combine shows a different but useful pattern: its gain is slightly smaller in absolute percentage, but more stable, with average speedup remaining around \(22\%\sim24\%\) across the tested hidden sizes and quantization modes. This stability is relevant in practice because decode workloads may run under different model shapes and deployment configurations.

Taken together, these results suggest that the proposed implementation improves decode latency across the tested configurations. The benefit is not tied to one hidden size or one quantization mode in this early experiment set.

\begin{table}[t]
\centering
\caption{Summary of decode-kernel latency reduction of the proposed implementation over the HCCL-enabled DeepEP baseline. Each row reports the mean latency across batch sizes \(\{16,32,64,80,128,144\}\), together with the average, minimum, and maximum speedup.}
\label{tab:decode-summary}
\resizebox{\linewidth}{!}{
\begin{tabular}{ll l ccccc}
\toprule
Operator & Hidden size & Quant mode & Baseline (\(\mu s\)) & Proposed (\(\mu s\)) & Avg. speedup (\%) & Min (\%) & Max (\%) \\
\midrule
Dispatch & 4096 & non-quant & 94.95  & 73.81  & 20.18 & 11.89 & 26.97 \\
Dispatch & 4096 & quant     & 88.62  & 59.46  & 31.02 & 18.71 & 39.91 \\
Dispatch & 7168 & non-quant & 124.73 & 104.48 & 14.83 & 8.90  & 19.16 \\
Dispatch & 7168 & quant     & 101.06 & 71.48  & 27.72 & 16.82 & 33.52 \\
\midrule
Combine  & 4096 & non-quant & 59.88  & 46.03  & 22.75 & 18.61 & 24.82 \\
Combine  & 4096 & quant     & 61.88  & 46.02  & 24.45 & 16.41 & 28.02 \\
Combine  & 7168 & non-quant & 89.76  & 70.23  & 22.43 & 19.79 & 28.34 \\
Combine  & 7168 & quant     & 91.41  & 69.80  & 24.34 & 20.89 & 28.79 \\
\bottomrule
\end{tabular}
}
\end{table}

The complete per-configuration decode results can be included in the appendix.

To further isolate the low-latency path, we evaluate two representative cases, namely a DeepSeek 3.1 case~\cite{deepseek2024v3} and a Qwen-235B case~\cite{qwen3}. Figure~\ref{fig:lowlatency-case-study} reports the corresponding dispatch and combine latency.

The case study shows gains on the low-latency path. In the DeepSeek 3.1 case, dispatch latency is reduced by \(23.3\%\) and combine latency by \(10.8\%\). In the Qwen-235B case, the dispatch gain reaches \(52.8\%\), while combine improves by \(18.3\%\).

This asymmetry between dispatch and combine is consistent with the structure of the low-latency kernels. Dispatch benefits most when direct token placement replaces a more explicit relay-oriented path, especially in cases where token redistribution is sensitive to coordination overhead. Combine also benefits from direct remote reading, but its relative gain depends more strongly on how much of the end-to-end kernel time is already occupied by weighted reduction. The two cases therefore suggest that the magnitude of the gain is model- and stage-dependent, while the direction of the gain remains positive in the tested settings.

\subsection{End-to-End Serving Performance}

Kernel-level improvements are most useful when they translate to serving-level behavior. We therefore evaluate end-to-end inference on a representative DeepSeek 3.1 serving scenario and report both \ac{TTFT} and \ac{TPOT}.

\begin{figure}[!tbp]
    \centering
    \includegraphics[width=0.8\linewidth]{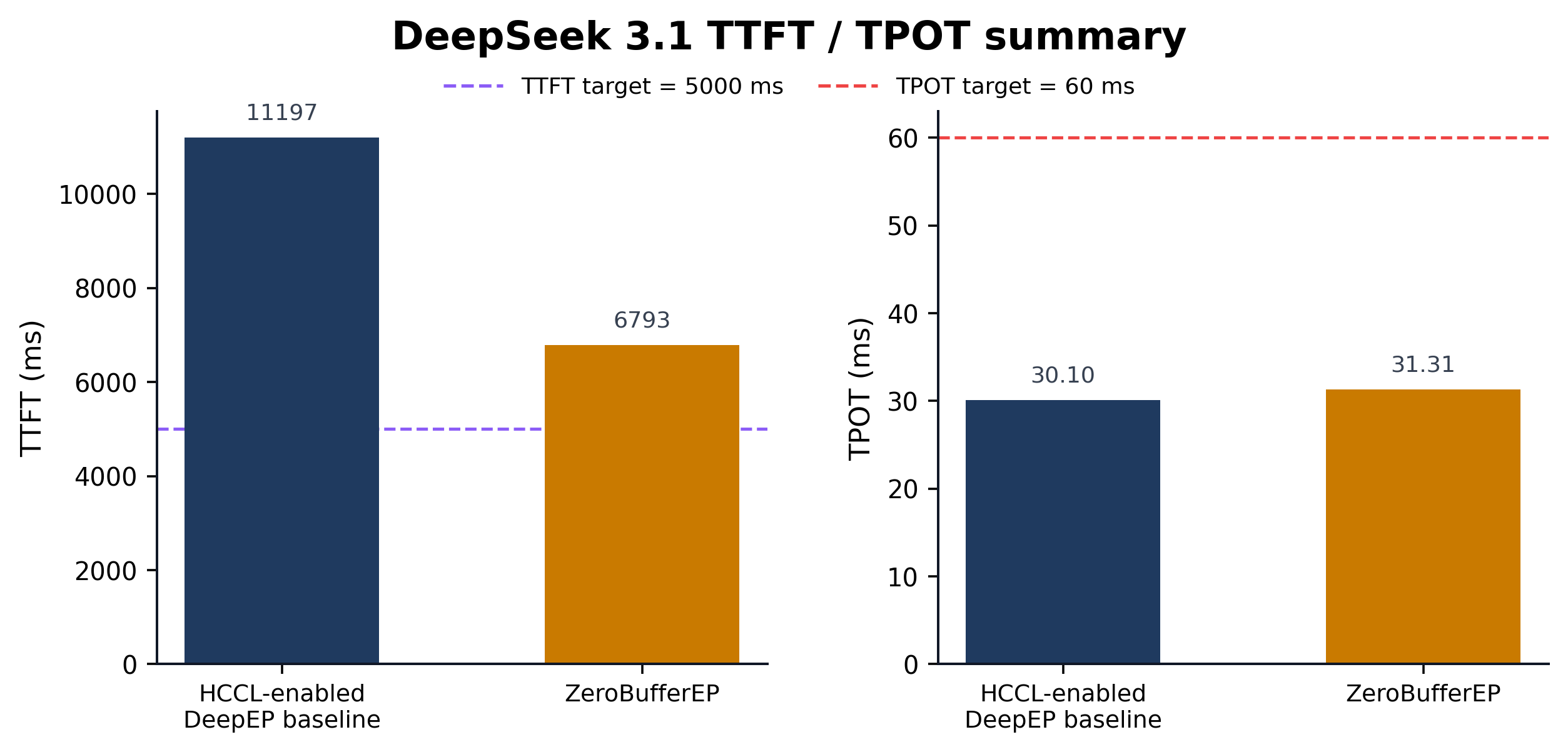}
    \caption{End-to-end \ac{TTFT} and \ac{TPOT} comparison on a DeepSeek 3.1 serving scenario.}
    \label{fig:ttft-tpot-summary}
\end{figure}

Figure~\ref{fig:ttft-tpot-summary} shows that the proposed implementation reduces \ac{TTFT} while leaving \ac{TPOT} in a similar range.

More specifically, \ac{TTFT} decreases from \(11197\) ms in the HCCL-enabled DeepEP baseline to \(6793\) ms, whereas \ac{TPOT} changes from \(30.10\) ms to \(31.31\) ms and remains below the target threshold shown in the figure.

This result suggests that the communication gain is mainly reflected in the prompt-processing stage, which is more exposed to dispatch/combine overhead and transient \ac{HBM} pressure. At the same time, the steady-state token generation speed remains within the target range in this scenario.

\subsection{Scheduling-Space Expansion}

\begin{figure}[!tbp]
    \centering
    \includegraphics[width=0.82\linewidth]{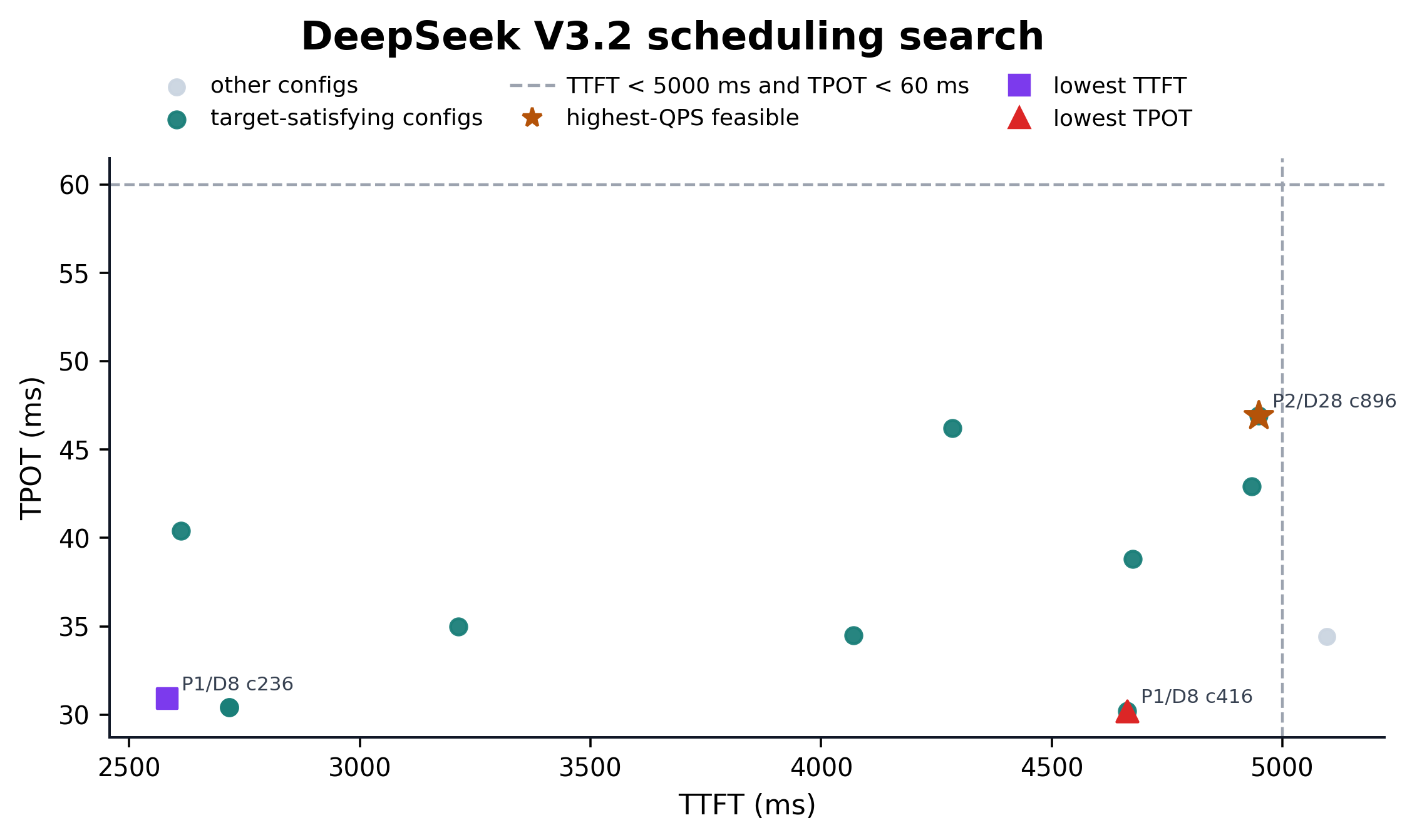}
    \caption{Scheduling search on a DeepSeek 3.2 serving scenario with respect to the targets \(\mathrm{TTFT}<5000\) ms and \(\mathrm{TPOT}<60\) ms.}
    \label{fig:dsv32-scheduling-search}
\end{figure}

Finally, we study a practical scheduling question on a DeepSeek 3.2 serving scenario~\cite{deepseek_v32}. Figure~\ref{fig:dsv32-scheduling-search} plots the scheduling configurations in the \ac{TTFT}--\ac{TPOT} plane under the target constraints \(\mathrm{TTFT}<5000\) ms and \(\mathrm{TPOT}<60\) ms.

The plot indicates that the proposed implementation enlarges the feasible operating region in this scenario. Several configurations satisfy both targets simultaneously, and the best throughput-feasible point lies close to the constraint boundary while still remaining inside the acceptable region. The figure also highlights a practical trade-off: the configuration with the lowest \ac{TTFT} is not necessarily the one with the lowest \ac{TPOT}. This means that improved communication efficiency is useful not only because it lowers latency in one fixed setup, but also because it gives the scheduler more room to choose among different operating points.

From a systems perspective, this complements the kernel-level results. The value of the proposed implementation is not only that it reduces individual dispatch/combine latencies, but also that it can make more serving configurations feasible under joint latency constraints.

\subsection{Summary of Findings}

Across the early experiments in this section, three observations are consistent.

First, the proposed implementation reduces both dispatch and combine latency in the tested prefill and decode settings. The gain appears in both the normal-kernel and low-latency paths, suggesting that the direct-placement and direct-read rule is not tied to a single execution regime.

Second, the gain tends to be stronger in settings where relay and reordering overhead are more exposed, especially decode dispatch and large-token prefill.

Third, the kernel-level benefit carries over to the tested serving scenarios. The proposed implementation reduces \ac{TTFT}, keeps \ac{TPOT} within the target range in the reported case, and enlarges the feasible scheduling space under the evaluated serving constraints.

Taken together, these results provide early evidence that reducing buffer-centric relay around expert routing can improve \ac{MoE} inference efficiency not only in isolated kernels, but also in end-to-end serving. Broader model coverage, more detailed memory-footprint measurements, and a more systematic serving benchmark are left for future versions of this report.
\section{Conclusion}

In this report, we presented a relay-buffer-free communication design for \ac{MoE} inference acceleration on Ascend systems. Rather than treating \ac{MoE} communication as a conventional relay-and-restore procedure, the proposed design reorganizes dispatch and combine around direct placement into destination expert windows and direct reading from remote expert windows. Built on top of globally pooled \ac{HBM} exposed through symmetric-memory allocation and shmem-style remote access, this design reduces explicit intermediate relay and reordering buffers around expert execution.

We further instantiated this design as two schedules tailored to the two main phases of \ac{MoE} inference, following the practical prefill/decode split used in DeepEP-style \ac{MoE} communication: a prefill schedule with richer planning state and a decode schedule with a more compact communication procedure. Although these two schedules differ in their degree of explicit planning and coordination, both follow the same underlying implementation rule: dispatch places token payloads directly into expert windows, while combine restores token-aligned outputs through direct remote reading and local weighted reduction.

Our early evaluation indicates that this design improves \ac{MoE} communication efficiency across both kernel-level and serving-level scenarios. At the kernel level, the proposed implementation reduces dispatch and combine latency in the tested prefill and decode settings. At the serving level, it reduces \ac{TTFT} while keeping \ac{TPOT} within the target range in the reported scenario, and it enlarges the feasible scheduling space under practical latency constraints.

More broadly, this report points to the importance of optimizing not only collective transfer bandwidth, but also the buffer-centric data movement around expert routing. For large-scale sparse serving systems, how routing results are transformed into communication state, how payloads are placed into expert windows, and how expert outputs are restored into token-aligned results can have a direct impact on both kernel-level communication cost and end-to-end serving behavior.

\begin{ack}
The authors would like to thank the colleagues and collaborators who contributed to the development, discussion, and evaluation of relay-buffer-free communication and \ac{MoE} inference acceleration on Ascend systems.
\end{ack}

\bibliography{biblio}







\end{document}